\pdfoutput=1 
\documentclass[letterpaper,twocolumn,10pt]{article}
\usepackage{usenix2025_SOUPS}
\AtBeginDocument{%
  \providecommand\BibTeX{{%
    Bib\TeX}}}
\usepackage{tabularx}
\usepackage{float}
\usepackage{placeins}
\usepackage{graphicx}
\usepackage{booktabs}
\usepackage{array}
\usepackage{subcaption}
\usepackage{fancybox}
\usepackage{multicol}
\usepackage[utf8]{inputenc}
\usepackage{tikz}
\usepackage{wrapfig}
\usepackage{placeins}
\usepackage{comment}
\usepackage{enumitem} 
\usetikzlibrary{shapes.geometric, arrows, calc, positioning, fit, arrows.meta}

\def\BibTeX{{\rm B\kern-.05em{\sc i\kern-.025em b}\kern-.08em
    T\kern-.1667em\lower.7ex\hbox{E}\kern-.125emX}}

\usepackage{caption}
\usepackage{subcaption}
\tikzstyle{startstop} = [rectangle, rounded corners, minimum width=3cm, minimum height=1cm,text centered, draw=black, fill=red!30]
\tikzstyle{process} = [rectangle, minimum width=3cm, minimum height=1cm, text centered, draw=black, fill=blue!30]
\tikzstyle{arrow} = [thick,->,>=stealth]

\tikzset{
    startstop/.style = {rectangle, rounded corners, minimum width=3cm, minimum height=1cm, text centered, draw=black, fill=red!30},
    process/.style = {rectangle, minimum width=3cm, minimum height=1cm, text centered, draw=black, fill=blue!30},
    branch/.style = {rectangle, minimum width=3cm, minimum height=1cm, text centered, draw=black, fill=green!30},
    arrow/.style = {thick,->,>=stealth}
}
\tikzset{
  startstop/.style = {rectangle, rounded corners, minimum width=3cm, minimum height=1cm, text centered, draw=black, fill=red!30},
  process/.style = {rectangle, minimum width=3cm, minimum height=1cm, text centered, draw=black, fill=blue!30},
  branch/.style = {rectangle, minimum width=3cm, minimum height=1cm, text centered, draw=black, fill=green!30},
  arrow/.style = {thick,->,>=stealth},
  box/.style = {rectangle, rounded corners, minimum width=2.2cm, minimum height=1cm, text width=2.2cm, text centered, draw=black, fill=softbox, align=center},
  caption/.style = {text width=2.2cm, align=center, font=\small\bfseries},
  research/.style = {text width=5cm, align=center, text=red!80!black, font=\bfseries\itshape},
  plus/.style = {circle, draw=black, thick, text centered, minimum size=0.6cm, font=\large\bfseries},
  mainarrow/.style = {thick, ->, >=stealth}
}

\definecolor{softbox}{HTML}{DCE4EE} 

\begin{document}
\date{}
\title{Toss If Perishable: An Ethnographic Study on Building Scenario-Based Training for Non-Perishable Skills}


\def\plainauthor{Francis Hahn, Spencer Cherry, Kumar Shashwat, Laura Araujo Buldrini, Daniel Lende, Xinming Ou}
\author{
{\rm Francis Hahn} \quad {\rm Spencer Cherry} \quad {\rm Kumar Shashwat}\\[0.2em]
{\rm Laura Araujo Buldrini} \quad {\rm Daniel Lende} \quad {\rm Xinming Ou}\\[0.6em]
University of South Florida\\[0.3em]
{\small\tt \{fhahn, sferrini, kshashwat, lauraaraujobuldrini, dlende, xou\}@usf.edu}
}
\maketitle
\begin{abstract}

Security Operations Centers (SOCs) often rely upon on-the-job training
focusing on the specific tools and procedures a SOC utilizes.  The complexity in
tooling can overshadow the underlying reasoning process, hindering an
analyst's ability to learn investigative skills.  We formulate the
concept of ``non-perishable knowledge'' which corresponds to
investigative thinking skills independent of tools.  We developed a
scenario-driven training method to understand whether such
non-perishable knowledge can be imparted through specially designed
incident scenarios, where trainees are presented with and solve
investigative challenges in a tool-agnostic manner.
We designed two
such scenarios based on real-world incidents. Human
subjects were recruited from a university's student body
for ethnographic study to understand how this training method is
received by the subjects and how they perform on such tasks. We collected
data from 20 hours of documented training with 25 trainees spread
across five sessions. Using grounded-theory, we analyzed the data and
uncovered factors that inhibit or promote learning of the
investigative thinking skills.  Our research combines scenario-based
training, ethnographic research, and technical analysis to examine how
to best train students in the reasoning skills that industry deems
vital to SOCs.

    \end{abstract}
    
\section{Introduction}
A Security Operations Center (SOC) is a critical component of an organization's cybersecurity posture. The workflow of a SOC is
comprised of many non-trivial tasks including but not limited to malware analysis, network packet inspection, and digital forensics. 
Understanding the full scope of a cyber incident and how indicators of compromise (IOCs) come together to form a threat timeline is
key to a successful SOC investigation. The core task of the SOC analyst is to understand how an incident occurred, prevent it from 
causing more loss, and document the events to allow rapid recognition and response to similar threats in the future. This type of 
understanding requires problem-solving skills not often taught in education settings.
\newline
\newline
On-the-job training is often used to train entry-level SOC analysts to fill the knowledge gap seen in those transitioning from academics to the workforce. Training in a SOC environment typically includes
learning the company-specific procedures to conduct an investigation and gaining familiarity with the tools utilized by the SOC.
They need to learn how to identify IOCs, collaborate with other analysts to get a holistic view of the data, and
understand the big picture or timeline of the incident. At the same time, they need to learn to write detailed post-mortem
reports 
to relay information to stakeholders and provide reference material to future analysts investigating similar incidents.
\newline
\newline
Work within SOCs is shaped by the playbook, a set of rules and procedures put together
specifically for various types of incidents. 
They prescribe a strict way of sequentially conducting an investigation of specific types of incidents and are often in the form of lists and/or
flowcharts that provide step-by-step actions an analyst needs to take for a given investigation. While playbooks are an essential
tool for controlling the consistency of analysts' performance,
they can be rigid and leave little room for creativity. This can hinder an analyst by setting
them into a standard way of operating when conducting an investigation that does not foster the use of reasoning skills.
While all SOC analysts refer to playbooks, senior analysts rely on them less often than entry-level analysts.
Analysts who
cannot move past the use of playbooks often find themselves stagnant in their roles, contributing to burnout~\cite{sundaramurthy2015human} and eventually leaving
the SOC. This creates a two-fold problem for SOCs: how to better train and retain their analysts
through instilling and promoting the problem-solving skills to successfully deal with cybersecurity incidents.
\\\indent
\newline
\newline
Through multiple consultations with members of a commercial SOC, we learned about
the environments they work in, the challenges they face at various levels of seniority, and how training is conducted. 
From those conversations 
emerged a notion of ``perishability'' of knowledge and skills. It is observed that some capabilities an analyst accrues
through on-the-job training ``perish'' over time, such as how to use features of specific SOC tools.
Those tool features can be overwhelming and hide the underlying reasoning process.
When analysts cease to use
those tools, e.g., by moving to a different role or a different SOC, those skills can be lost over time.
However, some ``core''
skills they learn on the job persist, such as how to initially approach an incident, identify critical data to further
investigation, and connect the dots among pieces of evidence. These skills tend to be ``tool-agnostic'' and reflect
an analyst's reasoning and problem solving capabilities. We use the word ``non-perishable'' to refer to
these core problem solving skills that enable analysts to effectively utilize any tools and their 
output to progress a cyber-incident investigation.
This raises the question: can the non-perishable knowledge and skills be imparted through a training method
that is designed to be tool-agnostic -- focusing solely on the investigative thinking process rather
than any specific features from SOC tools?
To this end we design a scenario-based training method we refer to as ``Mock-SOC,''
designed to emulate realistic investigations in a tool-agnostic way.
Trainees are presented with challenges to solve a cyber incident using
specially prepared data that obviates the need to use any specific SOC tools.
This allows trainees to hone in on how an analyst would problem-solve
a cyber incident without being burdened by the specifics of tool features,
and without having a standardized playbook to tell them how to proceed.
\\\indent
\newline
\newline
We recruited human subjects to participate in training research sessions based on two designed
Mock-SOC scenarios, and conducted an ethnographic study on the trainees while working
on the challenges. This
allowed us a unique perspective into how our trainees approached these cyber-incident investigations from a
reasoning skill-based approach compared to the common tool skill-based approaches. Through our study we observed
emergent and persistent themes such as what limitations the trainees faced, how comradery and collaboration factor into the success of an
investigation, how skill improvement can be achieved, and the rate at which it can occur. The results of our findings indicate that
scenario-based training can be created in such a way that focuses on the reasoning skills needed to be a successful SOC analyst and
formatted in a way which trainees enjoy and feel benefits their future careers.

\section{Methodology}
\label{sec:methodology}

We started by designing Mock-SOC scenarios that resemble a style of investigation that a SOC analyst would encounter in
an entry-level position. In the scenario design we prepared the data in a way that facilitates tool-agnostic investigations.
The training session is formatted to encourage the trainees to engage with the research team, providing ample opportunities
to collect ethnographic data.
We recruited 25 participants over three colleges in a public research university:
Engineering (which hosts the programs for Computer Science and Cybersecurity), Arts and Sciences,
and Business.
Two scenarios were designed; the first was used in the first three sessions with distinct participants
and the second was used in two later sessions. Four participants from scenario one's three sessions also participated
in one of the two sessions for scenario two.
We performed an ethnographic study on our trainees with 20 hours of documented observations
across the five sessions. 
Using grounded-theory and inductive analysis~\cite{bernard2016analyzing,thomas2006general},
we analyzed the various field notes to codify and identify themes observed and used this information to
understand how an entry-level analyst thinks when performing an investigation. Our findings were also used to refine future scenarios to better
understand what does and does not work in training the non-perishable skills.

\subsection{Data Collection}
To perform our ethnographic study, we needed to be able to analyze how the trainees performed in our Mock-SOC, both qualitatively and
quantitatively. These requirements led us to designing the scenario with opportunities to understand the thought process and rationale of
the trainees during their investigation. We used these opportunities to understand their decisions, how our training environment 
affected their approach, and if they find this type of training valuable. To facilitate this, we designed a story with critical points 
that were substantiated by evidence. Our design includes three main data sources:
\begin{enumerate}
    \item We required the trainees to make data requests directly to us instead of providing it to them up front. This gave us the
    opportunity to gain insight into their train-of-thought. We are able to question them on their thought process that led to the
    uncovering of the investigation's overall story bit-by-bit. By giving them a story to uncover, which encourages them to engage with us
    to progress, we get the opportunity to have conversations with them during their investigation. They are forced to engage with us
    instead of being engaged in technical tasks where documentation may have fulfilled any inquiries they may have had. 
    \item We required them to provide a written report on their findings, reasoning, and remediation. This allowed us to gain explicit
    detail on the non-perishable skills they used and the reasons behind their actions to understand the tacit knowledge used to reach
    their conclusion. We approached this data quantitatively and qualitatively by applying success metrics and making our own inferences
    to evaluate their performance and better understand how much they gained from working on the scenario.
    \item We ended every session with a round-table discussion to gather the opinions and a posteriori thoughts of the trainees. We 
    alternated in clockwise and counter-clockwise sequence asking questions pertaining to how they felt about certain aspects of the
    scenario, how they approached certain aspects of the investigations, and how they perceived the scenario-based approach
    compared to their learning
    experience in education or other training environments. This data provided us with further insights into how participants
    approached the training challenges after the fact and while not directly engaged in a time-sensitive task. We also used
    this data to evaluate the quality of our scenario and to understand how to improve the design for future ones.
\end{enumerate}

\subsection{Recruitment}
\begin{figure*}[h]
    \centering
    \includegraphics[width=\textwidth]{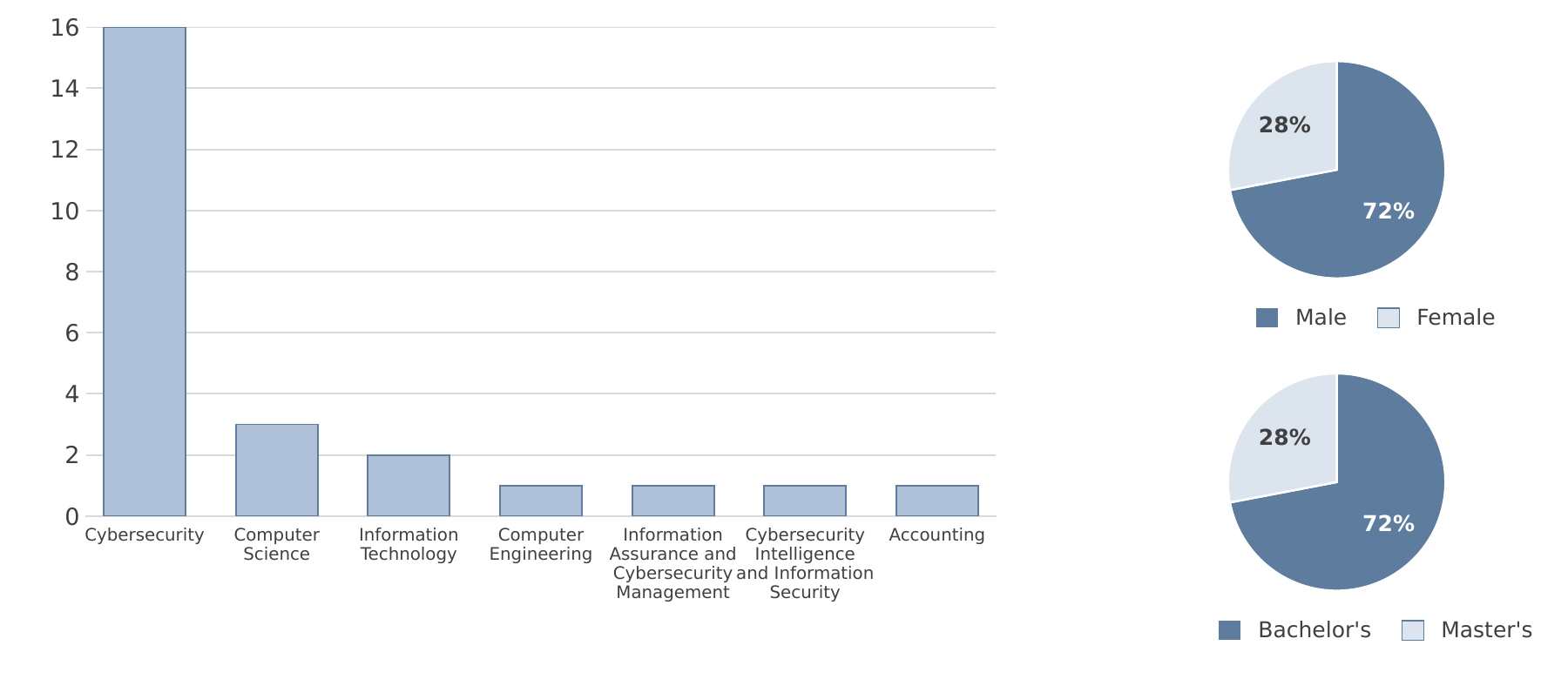} 
    \caption{Demographics for Participant Recruitment}
    \label{fig:demographics}
\end{figure*}

We recruit participants who self-identify as possessing basic knowledge in cybersecurity. Information about the study was sent
via emails and posted as flyers across campus. We focus the recruitment from student bodies in cybersecurity degree programs
across three colleges: Engineering, Arts and Sciences, and Business, as well as student cybersecurity clubs and competition teams.
After filling out informed consent, participants 
received a survey which asked them questions regarding what degree they were seeking, the current progress of their degree, what
cybersecurity courses 
they have taken, if they have ever had relevant industry experience, and what extra-curricular cybersecurity activities they participate in.
The aim of this was 
to obtain an adequate pool of applicants that are most similar to an entry-level SOC analyst. For our selection we chose students who
had completed courses which provided technical experience in cybersecurity, had relevant certifications, internship experience, or extra-curricular
experiences in cybersecurity. We had not considered any exclusionary criteria. Using our criteria we selected 25 of over 60 students who 
applied to participate in the research and placed them into 5 groups. Three groups were used for scenario 1 and two groups were used for 
scenario 2. Students who participated in scenario 1 were welcomed, encouraged, and prioritized to participate in scenario 2 sessions. Our 
reason for requesting repeat participation was to gather information on how a participant performs with an understanding of the format, 
but no understanding of the incident to begin observing longitudinal considerations. The demographics from our recruitment campaign are 
shown in Figure~\ref{fig:demographics}.

\section{Scenario Design}
\label{sec:The Scenarios}

Our approach to designing an effective scenario-based training required us to consider what it means to learn and the outcomes of genuine
engagement in institutional training programs. We considered aspects such as, when in an environment designed to teach specific
skills or concepts, some things are learned directly while others are learned indirectly~\cite{blum2019don}. In our scenario we consider
what is directly learned to be explicitly known skills that are documented and externalized by experts or educators, whereas what is
indirectly learned to be the skills or knowledge gained when one internalizes and discusses what they've learned 
directly~\cite{nonaka1996theory}, often through the work itself. This led us to three questions: (1) what skills do not transfer from one organization to the next,
(2) what amount of directly learned knowledge is necessary to induce the formation of tacit knowledge~\cite{Polanyi} -- knowledge possessed by
individuals but yet to be articulated, and (3) how can we analyze or gauge
the amount of tacit knowledge the trainees learned from our scenarios.
It is possible that a substantial amount of non-perishable knowledge our training method aims at imparting
is also tacit knowledge.
An example of such knowledge is reflected by SOC analysts' thinking process to extract IOCs from data in their investigations.
It is hard to find documented descriptions of how to perform this task, whereas skilled analysts learn them by doing the job.
Such skills bear the important hallmark of non-perishable skills, in that an individual can bring them from one job to the next.
To compare, knowledge and skills dependent on an organization, such as
how to use specific tools provided in a work environment, may not readily
transfer to a different work environment, and may fade away over time
after analysts cease to use them.
\newline
\newline
When designing our scenario-based training we drew on three primary data sources: (1) the expertise of SOC practitioners, (2) case studies, 
and (3) the curriculum for a university's cybersecurity program. From our discussions with practitioners, we identified concepts which help the more senior SOC analysts, or those who have made a career out of this line of work, perform more effectively. We translated these
concepts into skill sets or a set of experiences we wanted the trainees to learn. We then identified core issues entry-level SOC analysts
face such as repetitious tasks, reliance on playbooks, and a high pressure to learn job critical procedures on-the-fly. We also
looked at a university's curriculum to identify what could be used as an effective medium for deploying our scenario-based training. This
led us to using Microsoft's Windows Active Directory service because of its popularity among organizations, low saturation in typical curriculum material, and various security design flaws.
\newline
\newline
Our design first teaches the trainees a base level of knowledge for the concepts they will encounter. The trainees then undergo a
supervised investigation with minimal assistance and are periodically asked for their thoughts on the scenario and their approach to the
investigation. We conclude the scenario training by asking the trainees for a report that details their findings, remediation
suggestions, and thought processes. The intent of the reporting is to gauge the tacit knowledge they've acquired by having them make their
thoughts and approach methods explicit. Each scenario is intended to provide the trainees with information to serve as an initial starting 
point to work off of. To start the scenarios, each trainee is provided a ``service ticket'' that describes context on the organization and
the assumed attack they will be investigating. They are also provided a network diagram to help visualize the environment where the attack 
occurred, and an employee inventory and access policy that provides details on the employees, which machines they have access to, the 
department they work in, and which file shares they have permission to access.
\newline
\newline
To facilitate our training we built a small computer network consisting of 15 virtual machines, running a combination of Windows 10 Pro, 
Ubuntu 16.04, and Windows Server 2019. Using this network, we set up an Active Directory service with several accounts to mimic a small-scale
enterprise with employees. We then performed two separate chains of attack to create two distinct scenarios. Scenario 1 was  
a data breach due to a golden ticket attack \cite{duckwall2014kerberos}; scenario 2 was a ransomware attack. We collected data from
this network while conducting the attack scenarios 
and pre-processed the data to alleviate the need to use any specific SOC tools during the
investigation. 
Pre-processing was performed to condense the data into a more manageable format for the scenario, such as providing no more than 5 fields 
from any type of data and considering how human readable it was for participants.
For pre-processing we wrote various python scripts using known libraries to ingest and output the data to match our specifications.
This allowed us to administer the scenarios to a group of individuals where the sole focus was to think critically and
reason their way through a realistic cyber incident. 
\newline
\newline
In scenario 1, data requests were made by signaling to a member of the research team. The research team member then asked the trainee their reasoning and rationale for the data request and provided the trainee with the data if available. If the data was not available, the 
research team made slight suggestions, if the data they originally asked for was close to what was available, otherwise we informed them 
that their data request was unavailable. The intention is for the trainees to take the starting information, analyze its contents, and
use abductive reasoning to request additional data. In doing so, they are engaged in using both critical thinking and reasoning skills to
formulate how they believe the cyber incident occurred. This was changed slightly in scenario 2 by utilizing a data request ticketing 
system. While the methodology remained the same, the ticketing system eliminated face-to-face interactions with regards to data requests. 

\subsection{Scenario 1}

\begin{figure*}[h]
    \centering
    \begin{tikzpicture}[node distance=1.8cm, scale=0.9, transform shape] 
    \node (box1) [box] {Compromised\\ Employee Account};
    \node (box2) [box, right=1.8cm of box1] {Phishing Attack\\ Sent via Slack};
    \node (box3) [box, right=1.8cm of box2] {Compromised\\ Admin Account};
    \node (box4) [box, right=1.8cm of box3] {Golden Ticket\\ Attack};
    \node (box5) [box, right=1.8cm of box4] {Exfiltrate\\ Sensitive Data};

    \node[dashed, draw=black, thick, rounded corners, fit={(box1) (box2) (box3) (box4) (box5)}, inner sep=0.3cm] (dashedbox) {};
    
    \draw [mainarrow] (box1) -- (box2);
    \draw [mainarrow] (box2) -- (box3);
    \draw [mainarrow] (box3) -- (box4);
    \draw [mainarrow] (box4) -- (box5);
    
    \end{tikzpicture}
    \caption{Scenario 1 Cyber Incident Flowchart}
    \label{fig:scenario1}
\end{figure*}
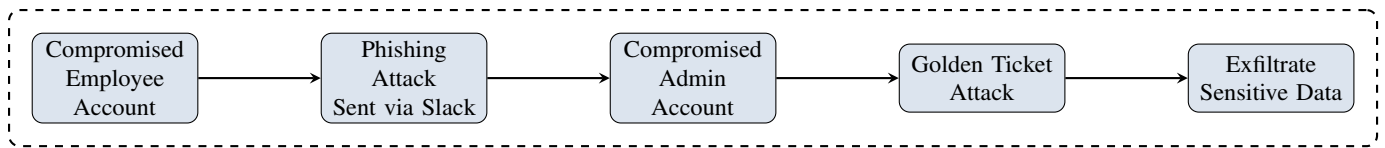

In scenario 1, the network was configured to have specific access control policies that segregated users to their own machines and file 
shares. Once setup, we performed a golden ticket attack on the network with the intention to model key elements exhibited during a similar 
real-world attack with the understanding that certain aspects needed to be relaxed given our limitations and control over the network. We 
enabled as many relevant logging features as possible to ensure the data we collected was rich in features indicative of the attack. We 
mimic an attacker as closely as we could, taking precautions to avoid the Windows malware detection and firewall rejection systems. We 
used various self-made powershell scripts, repositories, and obfuscation tools to perform the extraction and injection of Kerberos tickets. Using our 
tools we performed the necessary techniques to run mimikatz -- a popular tool for interrogating Active Directory and performing various attacks
against it, within the Active Directory managed network. This allowed the attacker to extract 
Kerberos tickets, forge their own tickets, and gain remote access to a file shared from a user computer that should not have permission to 
do so.
\newline
\newline
We designed this scenario with the following indicators of compromise in mind:
\begin{enumerate}
\item There were two phishing attacks (one implicit and one explicit) which gave the attacker access to the company network - a user machine and an administrator account with domain controller (DC) access privileges.
\item 
We used powershell scripts to embolden the powershell service to add a level of authenticity to our scenario data by disabling specific security features such as malware detection to evade signature-based detection.
\item Powershell was used to directly download files.
\item Mimikatz was used to extract and inject Kerberos tickets.
\item There was improper access to a file shared by a user not assigned access permissions. 
\end{enumerate}
These IOCs were provided in a collection of data sources which were pre-processed to alleviate the need to perform any technical actions
to obtain the data. This included the Windows event logs for logons to the DC and workstations, process logs for 
the DC and workstations, Kerberos requests to the DC, and a text log of the organization's slack 
channels. Figure~\ref{fig:scenario1} illustrates the attacker's actions in Scenario 1.
\newline
\newline
The learning goal of scenario 1 is for the trainees to use inductive reasoning to understand the attack path. The golden ticket attack is 
an exploit on access control circumvention and requires the trainee to understand the access control policies implemented. The
technical aspects of how Kerberos and the golden ticket attack work are important to the story but key story points or indicators of 
compromise come from recognizing improper pairings from the provided access policy map. The trainees need to request the proper logon 
logs to be able to recognize where the access control policies are not being properly enforced. By reasoning their way through the data, the objective is for the trainee to conclude that the data breach occurred by looking at the evidence.

\subsection{Scenario 2}

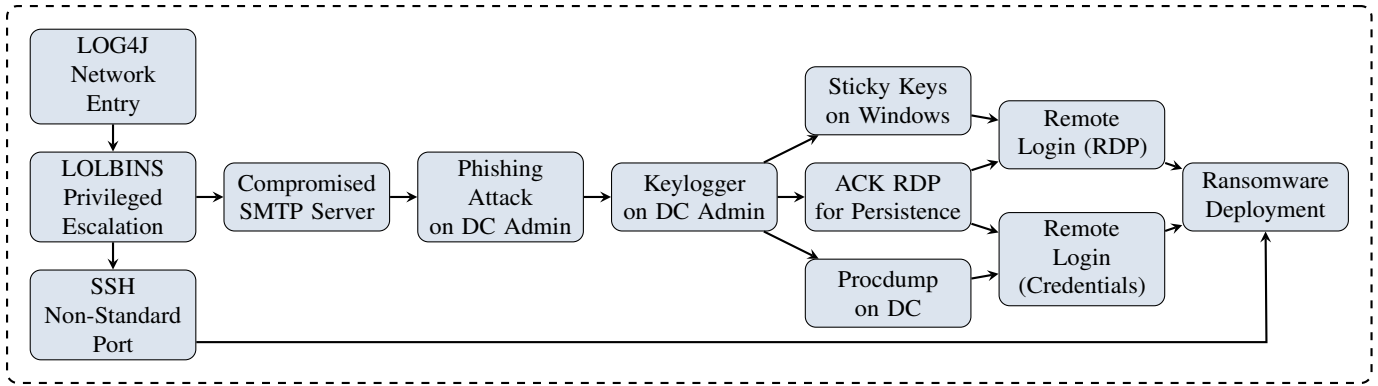
\begin{figure*}[h]
    \centering
    \begin{tikzpicture}[node distance=1cm, scale=0.9, transform shape] 
    
    \node (box1) [box] {LOG4J Network\\ Entry};
    \node (box2) [box, below=.4cm of box1] {LOLBINS Privileged\\  Escalation};
    \node (box3) [box, below=.4cm of box2] {SSH\\ Non-Standard Port};
    \node (box4) [box, right=.4cm of box2]  {Compromised\\ SMTP Server};
    \node (box5) [box, right=.4cm of box4] {Phishing Attack\\ on DC Admin};
    \node (box6) [box, right=.4cm of box5] {Keylogger\\ on DC Admin};
    \node (box7) [box, right=.4cm of box6] {ACK RDP\\ for Persistence};
    \node (box8) [box, above=.4cm of box7] {Sticky Keys\\ on Windows};
    \node (box9) [box, below=.4cm of box7] {Procdump\\ on DC};
    \node (box10) [box, right=.4cm of box8, yshift=-.5cm] {Remote\\ Login (RDP)};
    \node (box11) [box, right=.4cm of box9, yshift=.5cm] {Remote\\ Login (Credentials)};
    \node (box12) [box, right=3.1cm of box7] {Ransomware\\ Deployment};

    \node[dashed, draw=black, thick, rounded corners, fit={(box1) (box2) (box3) (box4) (box5) (box6) (box7) (box8) (box9) (box10) (box11) (box12)}, inner sep=0.3cm] (dashedbox) {};
    
    \draw [mainarrow] (box1) -- (box2);
    \draw [mainarrow] (box2) -- (box3);
    \draw [mainarrow] (box2) -- (box4);
    \draw [mainarrow] (box4) -- (box5);
    \draw [mainarrow] (box5) -- (box6);
    \draw [mainarrow] (box6) -- (box7);
    \draw [mainarrow] (box6) -- (box8);
    \draw [mainarrow] (box6) -- (box9);
    \draw [mainarrow] (box8) -- (box10);
    \draw [mainarrow] (box7) -- (box10);
    \draw [mainarrow] (box7) -- (box11);
    \draw [mainarrow] (box9) -- (box11);
    \draw [mainarrow] (box10) -- (box12);
    \draw [mainarrow] (box11) -- (box12);
    \draw [mainarrow, to path={ -- ++(15.73cm,0) -- ++(0,0) -- (box12.south)}] ([yshift=-0.4cm]box3.east) to (box12);
    
    \end{tikzpicture}
    \caption{Scenario 2 Cyber Incident Flowchart}
    \label{fig:scenario2}
\end{figure*}

The design of scenario 2 reused many aspects of scenario 1, namely the network design with Active Directory management and the existence 
of a golden ticket attack, but these aspects were not the main focus of this scenario. After performing the data analysis of scenario 1 
and speaking with our first round of trainees, we gathered valuable information which we used to enrich and change scenario 2's 
design. The focus of this scenario is a ransomware attack on web services and Linux machines that prevents the webstore from serving 
customers. We reused the virtual machines from scenario 1 and added an nginx server hosting the webstore which runs Log4J and an SMTP mail 
server for managing the organization's internal mail services. We found, from our analysis of scenario 1 data, that many of the trainees 
identified our IOCs well within the allotted time and expressed a desire for analyzing more logs. We incorporated this feedback into scenario 
2 with the additional intention of training the participants on living-off-the-land (LOL) techniques and supply chain attacks. The details of the full 
attack chain can be seen in Figure~\ref{fig:scenario2}.
\newline
\newline
We designed this scenario with the following indicators of compromise in mind:
\begin{enumerate}
\item There was an exploit of log4j on the webstore page to gain shell access to the host machine.
\item LOLBINs\footnote{A repository of living off the land (binaries) techniques: \url{https://lolbas-project.github.io/}} were used to escalate privilege on the host machine.
\item A configuration file was created for starting a new system daemon on a non-default port for SSH access.
\item The SMTP server was compromised to perform a phishing attack using an internal email.
\item A sticky keys attack was performed to establish persistence.
\item Ransomware was deployed to prevent any access to the affected systems by members of the organization. 
\end{enumerate}
On top of many of the original data sources provided in scenario 1, scenario 2 also included files which showed the use of the command
``ps aux'', the logs for the SMTP server, the results outputting all currently running systemd services, the log4j logs, output from the 
system32 directory, a capture of powershell script being run to extract the NTDS directory, and the directory that extraction resided in.  
\newline
\newline
The attack chain for scenario 2 focused more on technical facts because we wanted the trainees to deduce, from the evidence, a chain of 
events. Similar to scenario 1, we initially provided the trainees with general context but expected them to request evidence data. The 
focus for scenario 2 challenged the trainees to look at each piece of evidence they requested to find a localized issue and identify or 
reason how the issue plays into the overall attack chain. There was no notion of direction provided to the trainees; it was up to them to 
deduce the storyline by recognizing the issues and putting them together.

\section{Study Design}
\label{sec:experimentation}

Each study session was designed to be a three-to-four-hour event consisting of a primer, the cyber-incident investigation, reporting, a solution 
debrief, and a round-table discussion at the end. The primer was designed to give the trainees a base-line understanding of the concepts that will 
be present in the investigation. The Mock-SOC cyber-incident investigation follows immediately after the primer, where the trainees are given the 
initial data to begin their investigation.
At the end of the investigation we ask each trainee to write a report explaining their findings and thought processes.
Then we provide a debrief solution explaining one of the possible investigation 
paths for the scenario. We then take a 15 minute break and conclude with the round-table discussion.
The primer takes about 20-30 minutes. The investigation takes about 1.5 hours for scenario 1 and two hours for scenario 2.
Debriefing takes about 15 minutes. The roundtable discussion takes about 30 minutes. Participants can take breaks
during the investigation phase of the study, although few did so. Each study session has five to six participants.
There are about the same number of research team members in the room as well.
\subsection{The Primer}
When designing the scenario-based training we had two major considerations: (1) not all of our trainees have the same courses, 
exposure to material, and experiences and (2) we aimed to understand how to bridge the gap between education and on-the-job training for 
entry-level SOC analysts. We found the primer addressed both of these issues by using a condensed lecture-style format. We created a 20 to 
30 minute presentation that covers key topics our scenario had been designed around. We do not tell the trainees explicitly the 
purpose of the primer, but we inform them it is used to ensure all of them have a base-line understanding of important cybersecurity 
concepts. The primer serves to provide a uniform understanding of the concepts we want them exposed to prior to performing the 
investigation while also acting as an additional implicit dataset to aid the trainees in their investigation. While the primer mimics 
a classroom lecture style, it is narrowly focused and followed up with an immediate practical application in the subsequent investigation. 

\subsection{Cyber-Incident Investigation}
The cyber-incident investigation is the core of the scenario-based training design and where the entry-level SOC analyst skills are 
learned and practiced. This is also where the research team collects the majority of our data. Throughout the investigation the research 
team interacts with and observes the trainees while they perform the investigation.
Utilizing anthropological approaches to participant observation and informal interviewing~\cite{bernard2017research},
the researchers engage in face-to-face
interactions, prompting the trainees with questions. Our goal is to understand the different stages of their investigation, how they 
are approaching the investigation, and asking them if they have any questions. 
By using open-ended questioning rather than directive prompts,
the research team is careful not to nudge the trainees into 
an investigation path they would not organically enter to ensure they are exercising critical thinking and reasoning. We want the 
relationship between the researchers and trainees to be mutually agreeable so that we can have a candid and open discussion with them. 
This approach allows us to gather how they are thinking and to help supply them with the necessary information to progress in a meaningful way.

\subsection{Solution Debrief}
The solution debrief presented trainees with the overall picture of the cybersecurity incident and guided them through how an investigation
might have unfolded. We provided the trainees with a timeline of events comprising the security incident and discussed where the evidence is for each IOC. 
This debrief was an important facet of our educational approach, as we wanted to give the trainees an opportunity to see how an
investigation may have been conducted and compare 
with what they have done. 
In addition, giving them a solution puts 
an idea of the complete investigation in their minds to consider concurrently with how they conducted the investigation on their own. This 
works to help prepare them for some of the questions we will ask during our round-table discussion.

\subsection{Round-Table Discussion}
We conclude the scenario with a round-table discussion, which provides the research team and the trainees an open forum to discuss the
experience. The research team aims to gather specific information during the round-table by direct questions such as:
\begin{itemize}
\item What was your biggest take-away?
\item Did you get stuck at any point in the investigation? If so how did you move forward?
\item What do you think you could have done better?
\item How did you approach the investigation?
\item Could you see yourself doing this kind of work full-time?
\item How is this experience similar to and different from your curriculum? 
\end{itemize}
The trainees are encouraged to respond however they like and as candidly as possible. Before the round-table we provide the trainees with 
a solution and 15 minute break. We believe this allows them to begin to reflect on the investigation given the solution, time 
to analyze events, and time to consider their actions after the fact as used in post-solution based learning~\cite{zulharby2024post}.
Our intention for this line of questioning is to gather any final thoughts and how they believe this scenario-based training fills
the education gap with a combination of direct and indirect evaluation from the trainees.


\section{Analysis and Results}
\label{sec:discussion}
Our data analysis started by collating the different types of data we obtained, including the open-ended interactions during the scenario, 
the written requests and queries about those requests, the after-action reports and how that compared to the overall set of problem-solving 
tasks, and the group debrief which permitted much more participant and researcher interaction.
We combined qualitative research using anthropological methods with cybersecurity-based assessments of participant performance.
We individually analyzed the collected data and then came together as a team to discuss and share our findings to uncover any 
overlap. 
Some of the major themes we observed which later became codes are as follows:
\raggedcolumns
\begin{multicols}{2}
    \raggedright
    \begin{itemize}
        \item \textbf{Critical Thinking}
        \begin{itemize}
            \item Detective
            \item Technical
            \item Big Picture Thinking
            \item Investigation Methodology
        \end{itemize}
        \item \textbf{Limitations}
        \begin{itemize}
            \item Self-Imposed
            \item Tooling
        \end{itemize}
    \end{itemize}
    
    \vfill
    \columnbreak
    
    \begin{itemize}
        \item \textbf{Comradery}
        \begin{itemize}
            \item Nudging
            \item Community
            \item Collaboration
            \item Feedback
        \end{itemize}
        \vspace{1.7\baselineskip}
        \item \textbf{Skill Improvement}
        \begin{itemize}
            \item Insight
            \item Behavior
            \item Data Request
            \item Exposure
        \end{itemize}
    \end{itemize}
\end{multicols}


Our analysis was guided by both inductive analysis based on the scenario data and matching that analysis with 
our consultations with industry professionals and familiarity with how cybersecurity education happens in university settings.
A major goal of this analysis was to identify non-technical characteristics of a successful SOC analyst and use these as a guideline to 
further analyze data through multiple rounds of our grounded-theory approach. 

\subsection{Scenario 1 Observations}
The following describes observations from the findings generated from our analysis of scenario 1 and the data collected from the three 
sessions. Key observations about the trainees include:
\begin{itemize}
    \item They experienced challenges understanding where to get started.
    \item They were favorable to the challenge of not being provided data up front.
    \item Making connections was important to problem solving.
    \item One trainee seeking an information technology degree responded to being interested in SOC work, stating \textit{\textbf{``I've never heard a SOC before this training and could see myself working in SOC''}}.
\end{itemize}
The following three subsections give more detailed information about each session. These reportings are formatted to show the overall 
performance of the trainees and are meant to show some individual or differential observations found between each cohort.

\subsubsection{\textbf{Session 1}}
\begin{itemize}
    \item 3 of 5 or 60\% of participants showed a level of full scope understanding of scenario.
    \item 2 of 5 or 40\% of participants performed all required tasks as directed.
    \item Trainees in this session had a handful of self-imposed limitations that were perceived from their training environment. 
    We observed and recorded them mentioning that \textit{\textbf{``I don't feel like I can use google, this feels like a course exam''}}, as if they were in the mindset 
    similar to that of a closed-book examination which caused them to limit the usage of resources available to them.
    
\end{itemize}
\subsubsection{\textbf{Session 2}}
\begin{itemize}
    \item 3 of 6 or 50\% of the participants showed a level of full scope understanding of scenario.
    \item 3 of 6 or 50\% of the participants performed all required tasks as directed.
    \item Trainees that did not complete only found the phishing link.
    \item Some trainees used various 3rd party tools (i.e. VirusTotal).
    \begin{itemize}[label={}]
        \item \textit{\textbf{``When I did a virus scan on the website it came
        back with warnings about how it was malicious, phishing and another service said it
        was malware. I used IPQS as well to check and it said that it was hosting malware.''}}
    \end{itemize}
    
\end{itemize}
\subsubsection{\textbf{Session 3}}
\begin{itemize}
    \item 1 of 5 or 20\% of participants showed a level of full scope understanding of scenario.
    \item 2 of 5 or 40\% of the participants performed all required tasks as directed.
    \item Most trainees focused on the timeline of the incident.
    \begin{itemize}[label={}]
        \item \textit{\textbf{``My question now is? What happened after Adam clicked on the slack link 3 months ago? Did he proceed to do something that exposed the company in that way?}}
    \end{itemize}
    \begin{itemize}[label={}]
        \item \textit{\textbf{...}}
    \end{itemize}
    \begin{itemize}[label={}]
        \item \textit{\textbf{From the additional information given, Adam clicked on the supposed phishing link 3 months before the security incident occurred which means he is most likely to the cause of the security event.''}}
    \end{itemize}
\end{itemize}
\subsubsection{What We Learned}
Focusing on what we learned is critical to our study and iterative design process. From scenario 1, we found that our trainees had 
limited experience in formal reporting and documenting their thoughts and reasoning. This created a challenge for our data analysis as we 
relied mostly on recorded statements from the trainees. The importance of reporting not only belongs to our ethnographic study, but to the 
success of a SOC analyst. Because documentation in a SOC environment is important to inform future investigators, a SOC analyst is only as 
valuable as their ability to articulate their findings. This finding led us to approach reporting for scenario 2 differently, however this 
provided little change. Moving forward, we believe we need to provide training or a structured template for reporting. 
\newline
Our trainees mentioned that the primer made it obvious as to what attack was occurring. By being able to identify which attack is likely 
to occur based on the training helps accelerate developing the storyline and provides them with a sense of the big picture to consider 
when conducting their investigation. This alleviated some guess work and gave them additional information to make inferences about what data they 
should request. The feedback on scenario 1 was unanimously favorable by the trainees. We were fortunate enough to have a trainee with prior on-the-job SOC training where they stated \textit{\textbf{``Although I have worked in a SOC before this scenario pushed me to think outside the box. My previous trainings were done in a way where everything was prepackaged and there was no guess work to be done.''}} 

\subsection{Scenario 2 Observations}
Our observations on scenario 2 were different than scenario 1 due to the design changes we made. Scenario 2 was much more complex and presented an increased challenge for the trainees. Because of this we did not perform the same on quantitative evaluations, as the metrics used for scenario 1 would result in all of the trainees being labeled as incomplete. We believe this to be an inaccurate representation of their investigative journey. Instead, we focused more on what reasoning skills they exhibited during their investigation and how their data requests, which were logged by our tool, directed their progress. Those who made it further along the attack chain tended to find the localized issues, while the others focused more on understanding why the incident happened.
\subsubsection{\textbf{Session 1}}
\begin{itemize}
    \item Trainees felt not being given data upfront and having to rely on requests helped them learn more.
    \item One trainee mentioned having \textit{\textbf{``a light-bulb moment''}} occurred when they began relating data to the primer concepts.
    \item Our new trainees felt the amount of information in scenario 2 for the time allotted was overwhelming, while our repeat trainees 
    mentioned that the familiarity made it feel manageable.
    \item Many trainees mentioned they learned to \textit{\textbf{``trust their gut instinct.''}}
\end{itemize}
\subsubsection{\textbf{Session 2}}
\begin{itemize}
    \item A trainee with experience doing red-team testing was asked what they thought about the scenario, their response was \textit{\textbf{``this is a different type of challenge to think about an investigation as a defender instead of the attacker''}}.
    \item A trainee mentioned they used backtracking as a way of reasoning for their investigation.
    \item A trainee who attended scenario 1 mentioned that they felt limited, which we observed as a self-imposed limitation, of not wanting to \textit{\textbf{``metagame''}} by using details from 
    primer because it made them feel \textit{\textbf{``as if I were cheating.''}}
    \item When asked about how they thought to approach the investigation, one trainee said \textit{\textbf{``I used my previous training in looking for debits and credits as a method for looking for discrepancies in the data.''}}
    \item All of the trainees requested and investigated Log4J and SMTP logs, spending most of the investigation looking for indicators there. 
    \begin{itemize}[label={}]
        \item \textit{\textbf{``Only logs available were Log4J, from 3 days, I focused searching the logs that are on the same day as the malicious process was created.''}}
    \end{itemize}
    \item 2 of 9 trainees requested service dumps and process dumps.
    \item 3 of 9 trainees requested the folder containing the NTDS dumps.
\end{itemize}
\subsubsection{What We Learned}
Scenario 2 was our first attempt at using our data analysis to refine the scenario itself. We found there was still an issue with how to 
receive proper and meaningful reporting data from the trainees. From this, we've observed the concept of reporting on a timeline of events 
and documenting a collection of one's own thought processes is not something inherently known or taught. We believe using a structured 
reporting template and including a brief lesson on reporting in our primer would be beneficial. The increase in complexity made it 
challenging for some trainees to grasp all of the concepts introduced in the primer and properly apply them in the allotted time. This 
scenario was also observed to have a lower amount of in-session collaboration compared to what was observed in scenario 1. Possible 
refinements would be to recommend collaboration at a certain point to be more representative of a SOC or have this as a 2-day training 
scenario to break up the learned material. This would allow us to expose additional data after each primer and give the trainees an 
opportunity to distill and form tacit knowledge on what they're learning in a more distributable manner. 
Despite the challenges faced the trainees exhibited the skills and investigation practices characteristic of successful SOC analysts and 
stated there was value in the challenge, noting particular growth once exposed to the solution debrief. This type of learning shows that 
by engaging them in a challenging style of learning they begin to synthesize their own ways of approaching and forming tacit knowledge. 
Better ways of extracting the tacit knowledge post solution debrief would be of great value in future scenarios. 

\subsection{Critical Thinking}
Beyond the specifics of the cyberincident, we find that trainees implicitly draw on inductive, deductive, and abductive reasoning to analyze 
the data and formulate the incident details. We observed that the trainees at 
times acted in the way that a detective would by searching for or through evidence to support their ideas and finding someone to blame. 
While we designed this scenario-based training to be non-technical in deployment, there is an implicit need to understand and utilize 
technical knowledge. We observed the trainees used their technical skills to make sense of the data and consider what happened and how. 
This can be directly seen from recorded dialogue using our chat tool.
\begin{itemize}[label={}]
    \item \textit{\textbf{``...i think getting access to full data , gives me chance to analyze what we have in there and where to start, cause ransomware can be tapped by malicious emails or any customers data that might be saved online.''}}
\end{itemize}

\begin{itemize}[label={}]
    \item \textit{\textbf{``I would like to request the APACHE server logs, to look for unusual requests that could be code injections.''}}
\end{itemize}
\begin{itemize}[label={}]
    \item \textit{\textbf{``Since payment processing is down, I would like to request data from any chat logs which might have suspicious links or emails from the Sales or accounting group.''}}
\end{itemize}
The trainees used their technical knowledge to guide their investigation by understanding how malicious actions can be deployed.
It can be seen the trainee had a variety of approaches which exemplifies how individual experiences and thought processes contribute to the success of 
cyber investigations. We found that the most challenging aspect of reasoning through an investigation is big picture thinking, or the 
ability to deduce a timeline. This, however, is a critical aspect of a SOC analyst's job. The scenarios we supplied were non-trivial
and contained multiple IOCs that chained together. This required the trainees to find them all in order to truly understand the scope and 
events as they occurred. We noticed that all trainees were successful in their investigations, but varied in their ability to 
convey all of the facts, some focused on IOCs and some focused on the timeline. We believe this singular approach to the investigation
leads to variation in success contributes to trainees getting on their path of investigation, otherwise known as 
rabbit-holes.
\begin{itemize}[label={}]
    \item \textit{\textbf{``...we need some information related to the hacked source, time stamp and the affected data. so that we can try to find out what and how it happened.''}}
\end{itemize}

\subsection{Limitations}
Throughout our discussions with the trainees we noticed certain factors which we categorized as limitations, either tooling-related or 
self-imposed. The tooling-related limitation shares characteristics of technical themes but is unique to how they inhibited the trainees 
progress. Two major observations of tooling-related limitations were noted.
\begin{enumerate}
    \item The use of the signature-based detection tool, ``VirusTotal'', allowed some trainees to explicitly identify malicious files and links. 
    This contributed to the overall content of their report and allowed them to avoid going down rabbit holes by giving concrete proof to their 
    findings. Those without the knowledge of this type of tool were unable to validate their findings with certainty which led to observing 
    uncertainty in their investigation approach and reporting as seen in the following 
    \begin{itemize}[label={}]
        \item \textit{\textbf{``From here didn't exactly know where to go i had some information I found what seems to be the malicious program but really needed to find out how this program got to Adams machine''}}
    \end{itemize}
    
    \item  We introduced a data request ticketing system  for scenario 2. Our observation of this limitation has noise associated with it 
    due to a change in the investigation's content and a nearly unique group of trainees between each scenario. However, we believe that the 
    tool, while simple to use, added a layer of tooling usage which inhibited our ability to have open discussions with trainees as they 
    requested data. We attempted to have the same type of dialogues but as with any form of text-based communications dimensionality 
    is naturally lost. We also found this 
    tool to impose a collaboration limitation on the trainees due to an observation. We noted that in the scenarios where we were having 
    open dialogues with some of the trainees others would overhear and learn from our discussions. This form of community-based learning 
    was lost with the introduction of the chat tool. For some trainees this also reduced the dialogue to just a data-request.

    \item The self-imposed limitation emerged through trainees sharing that they assumed the investigation was a closed-book assessment, which we believe 
    to be an artifact of their academic training. We believe this limitation extends into entry-level environments, as it's a learned behavior 
    that is shed as a trainee transitions from the academics to the workforce. This is an important factor when considering incorporating this 
    scenario design into an academic environment as this limitation could dilute the learning outcomes.
\end{enumerate}

\subsection{Skill Improvement}
Identifying themes relating to skill improvements because we consider this to be a success criterion. 
During the scenarios, we noticed that 
data requests were sometimes made due to information from our primer. Because of this observation, we explicitly asked the trainees if the 
primer has aided in their decision making. The trainees confirmed our hypothesis, leading us to regard this as an exposure and insight 
based skill improvement. To the best of our knowledge and based on our recruitment data, over 90\% of the trainees have not been 
introduced to the concepts present in the scenarios. This supported our hypothesis on whether exposing trainees to new content and having 
them reason through the scenario would increase the tacit knowledge they gained. Then, by having them request data and report on the 
results, we engaged them in making the tacit knowledge explicit. This cycle of knowledge acquisition resulted in a behavior change within 
the trainees towards that of a successful SOC analyst, as the trainees were able to think more critically and confidently.

\subsection{Comradery}
The theme of comradery is important but challenging to identify. From our conversations with SOC analysts we found there is a significant 
amount of collaboration between SOC teams and the individuals within those teams. However, our training scenario is a one-off event with 
people that have had minimal prior interactions. We noted that, across all scenarios, trainees were conversing with each other, the 
research team, and listening to open discussions about the scenario and its data. 

\begin{itemize}[label={}]
    \item \textit{\textbf{trainee - ``i also believe EventID: 4688 is very suspicious with the given context from above''}}
    \item \textit{\textbf{researcher - ``You may google event ids you think are suspicous to understand them better''}}
    \item \textit{\textbf{researcher - ``please continue to let us know your thought process and where this investigation takes you, this is helpful for our research.''}}
    \item \textit{\textbf{trainee - ``in the active directory dump, one of the files ntds.dit is a file known to be used as a password extraction vulnerability. I believe that using the administrative privileges, the attackers were able to gain access to user account information and hold their information for ransom''}}
    \item \textit{\textbf{trainee - ``I also believe that having access to the registry from the shared temp folder is a vulnerability that the attackers may be exploiting''}}
    \item \textit{\textbf{researcher - ``We do have some additional information on ntds, how might an attacker deploy malware?''}}
    \item \textit{\textbf{trainee - ``the sethcbk.exe process in the system32 dump looks to be suspicious and potentially used to a sticky keys attack''}}
    \item \textit{\textbf{trainee - ``any additional information would be very helpful''}}
    \item \textit{\textbf{trainee - ``it is possible that the attackers could be deploying the malware through the domain controller, as well as through the SMTP server''}}
\end{itemize}
While the trainees had no inherent trust or community coming into the sessions, 
we observed a sense of trust and community quickly form. We believe this comes from selecting trainees with similar backgrounds and interests. 
This observation is important because it is our belief that a sense of community is required for meaningful collaboration to occur. 
\newline
Even though SOC analysts are often part of a team, they have to work individually on tasks, similar to our scenarios.  Participants drew on the social 
resources available to them - the primer, interactions with the research team, overheard conversations, responses to requests - as part of how 
they problem solved.  They also explicitly drew on prior social experiences, whether those were security-related experiences and/or cybersecurity 
classes.  Our analyses highlight an interesting dilemma: Prior experience could both help and hinder.  If participants know a way to try to solve 
the scenario that was akin to a playbook, they often tried that.  This prior reliance on social experience could be limiting in regards to the 
technical demands of the scenario.  Yet it was precisely through seeing other progress and open-ended questioning that participants recognized that 
they were not proceeding at the same pace.  This challenge, combined with the generalized primer and open-ended questioning, often prompted participants 
to open up and start to problem solve in new ways.  This point - that social interaction shapes problem solving - is an important dimension that the 
Mock SOC has revealed.

\section{Ethics}

This human-subjects research was approved by the university IRB where the research was conducted, and all participants provided informed consent. As part of informed consent, participants received information on the benefits and risks of participating in the research before deciding to voluntarily participate in the research. Benefits included learning more about cybersecurity. Risks included potential violation of confidentiality and possible embarrassment, due to the nature of the training study where multiple subjects participate in one session. Both risks were considered low. During the data collection, participants were reminded that their participation was voluntary, and they were not required to answer questions and could exit the study without consequence. All data collected was handled in a secure manner and stored using a password protected repository where all researchers had enforced multi-factor authentication in-place. Anonymization was performed during data processing and maintained during write-up and publication.

\section{Limitations}
\label{sec:Limitations}
A largely influential limitation of this work is the lack of research done in the field of study. Studying the 
culture of SOCs and how to better train SOC analysts has been looked at in the past, but not extensively. 
The limited work done in this field gives us not much to build upon. The scarcity of research forces us to rely
heavily upon interdisciplinary collaboration. While in itself not a limitation, the difficulty in finding research
members of dual-disciplinary interest is. Another limitation we faced for this work is having short windows of opportunity to 
observe our trainees and interview them. This made it challenging for us to take our observations 
and gather deeper insight after-the-fact when questions arose.
\section{Related Work}
\label{sec:related}
It is clear that SOCs play an important role in all types and sizes of organizations \cite{kaliyaperumal2021evolution}. 
Despite this, SOC-specific research is limited \cite{kokulu2019matched}. Although the issues SOCs face are many, people 
tend to be the biggest hindrance, so we aim to address this critical issue. A comprehensive review of the work available 
shows that the human factor is a major problem \cite{nobles2022stress} \cite{chamkar2022human} of the SOC, including 
insufficient analyst training \cite{kokulu2019matched}. Prior work \cite{sundaramurthy2015human} investigated the idea 
of SOC burnout, based on the ``vicious cycle'' of the job. Drawing from economics, they proposed a solution where analysts 
were considered ``Human Capital'' who need training and support to effectively perform their duties. Our work builds upon 
this by taking the human capital model and applying it to our scenario-based training to improve the skills of trainees 
by putting them in an environment where they must think creatively in a low-stress and low-risk environment. We aim this 
study at higher education students because we believe they have the most similar competency to an entry-level analyst. 
While critical thinking is a key goal of higher education it is often neglected for discipline-specific knowledge \cite{bellaera2021critical}. 
We use this to focus the scenario-based training ideal outcome on teaching the critical thinking skills required for a SOC analyst. 
Our goal is that they walk away with new insight and knowledge regardless of their performance. Another work \cite{274475} planted
computer science students trained in anthropological methods in a software development team to observe and attempt to 
create a culture focused on secure coding practices. Our work builds upon this approach by observing the roles of SOC 
analysts of various career stages. We use discussion sessions to better understand the culture which, based on our observations, 
appears to be unchanged in over a decade. These discussions and observations assist us to build near authentic scenario-based 
training modules. Our goal is to design a framework that can be used by others to administer training or modify for their own 
use. This work leverages the methods found in research \cite{lende2023co} involving the co-creation of solutions alongside the 
analysts, the managers, the researchers, and the trainees. There is an observed research-practice gap in human-centered cybersecurity 
research \cite{haney2024towards_Bridge} \cite{haney2024towards_Integrating} between practitioners and researchers. This gap has 
been seen to reduce the effectiveness of research by a two-fold issue of a lack of interest by practitioners and a poor interaction 
timing by the researchers. Our work aims to document an approach to overcoming this problem through continuous engagement, allowing 
us to provide authentic experiences in our scenario-based training and provide usable documented training modules for SOC practitioners 
that are guided by their peers. In our work we consider the results or outputs of software-based tools as data \cite{lende2023software} 
and because we consider tooling skill sets to be perishable knowledge we separate the user, our trainee, from the need to produce the 
data themselves. Thus we alleviate the need to use the tools during an investigation and this allows us to provide our trainees a 
tooling-free investigation environment. Our work uses this method to enable the teaching of non-perishable skills because through our 
discussions with senior SOC analysts we have concluded these skills to be lacking in the more junior analysts. A SOC analyst armed with 
experience, technical skill, and the ability to critically think empowers them with the capability to think freely and objectively to 
generate meaningful paths for incident investigation.


\section{Conclusion and Future Work}
\label{sec:Conclusion and Future Work}
In this work we've performed an ethnographic study by creating a Mock-SOC scenario-training to conduct human-subject research. Using 
students from a university we had them participate in an investigation event where they learned and used various critical thinking and 
reasoning skills. This produced data that allowed our research to discover whether or not a scenario-based training can be built using our 
concept of non-perishable knowledge to emphasize the learning of the reasoning skills characteristic of a successful SOC analyst. Our 
future work will aim to further improve the scenario-based trainings using our data, conversations with SOC teams, and iterative design 
process. Some methods to improve our research are including more structured reporting to both train the subjects in the reporting
techniques and to gain more informative data on their thoughts. We also plan to approach the training scenario as a longitudinal study by 
attempting to recruit trainees who can consistently attend repeat scenarios. This will allow the research team to collect more consistent 
data on the effects of our refinements in a controlled setting. Our research aims to embed PhD students within a SOC to gain first-hand
experience of what working in a SOC is like to be able to better understand the stressors, tools, and culture to improve the authenticity
and effectiveness of training scenarios. In building higher quality scenarios we would attempt to extend this work to use actual SOC analysts
as trainees to gather what benefit it could be to actual professionals. We conclude this work acknowledging that such a scenario-based training 
can be built as evidenced by our trainees' favorable responses identified by our collected data and uncovered using various analysis methods.

\section{Acknowledgements}
\label{sec:acknowledgement}

We’d like to thank all of the SOC practitioners we spent time in discussions with for their help in this research and our corporate collaborator for making their team members available to us. We’d also like to thank our many trainee
volunteers for their time and effort to make this research possible. An additional thanks goes out to all of those who we held outside discussions with who helped influence the work at various points.

This work was partially supported by the National Science Foundation under award no. 2235102, and Office of Naval Research under award no. N00014-23-1-2538.

\bibliographystyle{unsrt}
\bibliography{usenix,anth}
\appendix
\section{Recruitment Survey}
\label{sec:recruitmnet survey}
\subsection*{Consent}
Our recruitment survey began with an IRB approved consent for minimal risk. This outlined the purpose of the study, the risks and benefits, and the right to withdraw at any time. 
Participants were informed that their responses would be used for research purposes only and that they could withdraw from the research at any point.

\subsection*{Demographics}
\begin{itemize}
    \item What is your Name?
    \item What is your email address?
    \item What is your gender?
    \item What is your race/ethnicity?
    \item What is your country of origin?
\end{itemize}

\subsection*{Education}
\begin{itemize}
    \item Are you currently enrolled at USF?
    \item What degree are you pursuing?
    \item What is your major?
    \item What is your expected graduation date?
    \item Please tell us about any cybersecurity courses or computing courses you have taken.
\end{itemize}

\subsection*{Computing Background}
\begin{itemize}
    \item Have you written a program before? If yes, which programming languages have you used?
    \item Please mention any cybersecurity tools that you have used. Eg: Wireshark
    \item Please tell us about any relevant experiences you have (e.g. job, projects, etc.)
    \item Upload Resume (optional)
\end{itemize}
    
\subsection*{Availability}
\begin{itemize}
    \item Which of the following sessions can you participate in? Check all that apply. Even if none works for you, please still complete the survey so we can invite you for future sessions.
\end{itemize}
\section{Post-Mortem Interview}
\label{sec:post-mortem interview}
\begin{itemize}
    \item What tools have you learned to use?
    \item What lessons have you taken from this experience?
    \item Do you feel this scenario-based training fills a gap in your education?
    \item What did you like most about the training?
    \item What did you dislike most about the training?
    \item Is there anything you would change about the training?
    \item If you could create something that would model this experience to share with others, what would it be?
    \item Has this scenario-based training caused you to consider differently any aspects of cybersecurity?
    \item Has this scenario-based training changed how you would consider approaching an investigation?
\end{itemize}

\end{document}